\documentclass[%
 reprint,%
 amssymb, amsmath,%
 aip,cha,%
]{revtex4-1}

\usepackage{docs}%
\usepackage{bm}%
\usepackage{graphicx}%
\usepackage{subfigure}%
\usepackage[colorlinks=true,linkcolor=blue]{hyperref}%
\expandafter\ifx\csname package@font\endcsname\relax\else
 \expandafter\expandafter
 \expandafter\usepackage
 \expandafter\expandafter
 \expandafter{\csname package@font\endcsname}%
\fi
\begin{document}

\title{Vector Dissipative Solitons in Graphene Mode Locked Fiber Lasers}%

\author{Han Zhang,$^{1}$ Dingyuan Tang,$^{1,*}$ Luming Zhao,$^{1}$ Qiaoliang Bao,$^2$ and Kian Ping Loh ,$^2$}%

\address{
$^1$School of Electrical and Electronic Engineering, Nanyang Technological University, Singapore 639798
\\
$^2$Department of Chemistry, National University of Singapore, Singapore 117543 \\
$^*$Corresponding author: http://www3.ntu.edu.sg/home2006/zhan0174/\\
}
\email{http://www3.ntu.edu.sg/home2006/zhan0174/}

\date{February 2010}%
\revised{\today}%
\maketitle

\tableofcontents

\section{Introduction}

Vector soliton operation of erbium-doped fiber lasers mode locked with atomic layer graphene was experimentally investigated. Either the polarization rotation or polarization locked vector dissipative solitons were experimentally obtained in a dispersion-managed cavity fiber laser with large net cavity dispersion, while in the anomalous dispersion cavity fiber laser, the phase locked NLSE solitons and induced NLSE soliton were experimentally observed. The vector soliton operation of the fiber lasers unambiguously confirms the polarization insensitive saturable absorption of the atomic layer graphene when the light is incident perpendicular to its 2D atomic layer.

\subsection{Soliton Fiber Laser}

Soliton operation of passively mode locked fiber lasers has been extensively investigated \cite{no1,no2,no3,no4}. Traditionally, soliton operation of fiber lasers was found on fiber lasers with net negative cavity dispersion, where due to the natural balance between the cavity dispersion and fiber nonlinear optical Kerr effect, a nonlinear transform-limited pulse known as optical soliton is automatically formed \cite{no5}. It has been shown that the dynamics of the formed soliton pulse is well governed by the nonlinear Schrodinger equation (NLSE), a paradigm equation describing the Hamiltonian solitons. Recently, it was further found that optical solitons could even be formed in a mode locked fiber laser operating in the net positive cavity dispersion regime, where no natural balance between the cavity dispersion and the fiber nonlinear optical Kerr effect exists \cite{no6}. The dynamics of the solitons formed in the normal dispersion cavity fiber lasers is governed by the Ginzburg-Landau equation (GLE). Solitons whose dynamics is described by the GLE are also known as dissipative solitons \cite{no7}. Dissipative solitons formed in the net positive dispersion cavity fiber lasers have the characteristics of steep spectral edges and large frequency chirp, which are distinct from those of a NLSE soliton formed in the net negative dispersion cavity fiber laser \cite{no3}. Moreover, a dissipative soliton has much larger pulse energy and broader pulse width than a NLSE soliton, which can be utilized for generating large energy ultrashort pulses from a mode locked fiber laser \cite{no8,no9,no10,no11,no12,no13,no14,no15}.
\subsection{Mode Locking Fiber Laser}

Conventionally, mode locking of a fiber laser is achieved with the nonlinear polarization rotation technique \cite{no6}. To implement the technique in a fiber laser a polarizer or a polarization discriminating component is inserted in the laser cavity. The incorporation of a non-fiber polarizing component not only results in extra cavity loss but also limits the polarization dynamics of the formed solitons, because the intracavity polarizer fixes the soliton polarization. Therefore, solitons formed in the mode locked fiber lasers are scalar solitons. A fiber laser can also be mode locked with polarization independent saturable absorbers, such as the semiconductor saturable absorption mirrors (SESAMs).  It has been shown that vector solitons, these are solitons with two coupled orthogonal polarization components, could be formed in a SESAM mode locked fiber laser. Depending on the net cavity dispersion, either the vector NLSE type of vector solitons or the dissipative vector solitons have been experimentally demonstrated \cite{no16}. The vector soliton operation of a fiber laser is attractive as it provides a test bed for the experimental investigation on the dynamics of solitons governed by the coupled NLSEs or the coupled GLEs.

\subsection{Graphene Mode Locked Vector Solitons}

In this contribution, we report that, apart from SESAMs, atomic layer graphene can also be used as a polarization independent saturable absorber for generating vector soliton in a fiber laser. Using atomic layer graphene based novel mode locker in an erbium-doped fiber laser, we have experimentally generated both the polarization rotation and the polarization locked vector NLSE solitons and dissipative vector solitons. Graphene is a single layer of carbon atoms arranged in a honeycomb lattice. It is so far the only known true 2-dimensional material. Graphene has a zero energy band-gap with a unique linear energy dispersion relation \cite{no17}. Since first isolated in 2004, graphene has created a revolution in condensed matter physics and been extensively investigated as the next generation material for nano-electronics applications. Previous studies have shown that graphene has a super broadband linear optical absorption, extending from visible to mid infrared \cite{no18}. Recently, we have further shown that graphene possess wavelength independent ultrashort saturable absorption, which can be exploited for the passive mode locking of fiber lasers \cite{no19,no20,no21}. Comparing with SESAMs, graphene as a saturable absorber for fiber laser mode locking has the following advantages: (i) controllable saturable absorption strength through controlling the number of graphene layers or chemical functionalization; (ii) super broadband saturable absorption; (iii) ultrafast saturation recovery time; (iv) easy to be fabricated.

 \begin{figure}
\centering
\includegraphics[width=8cm]{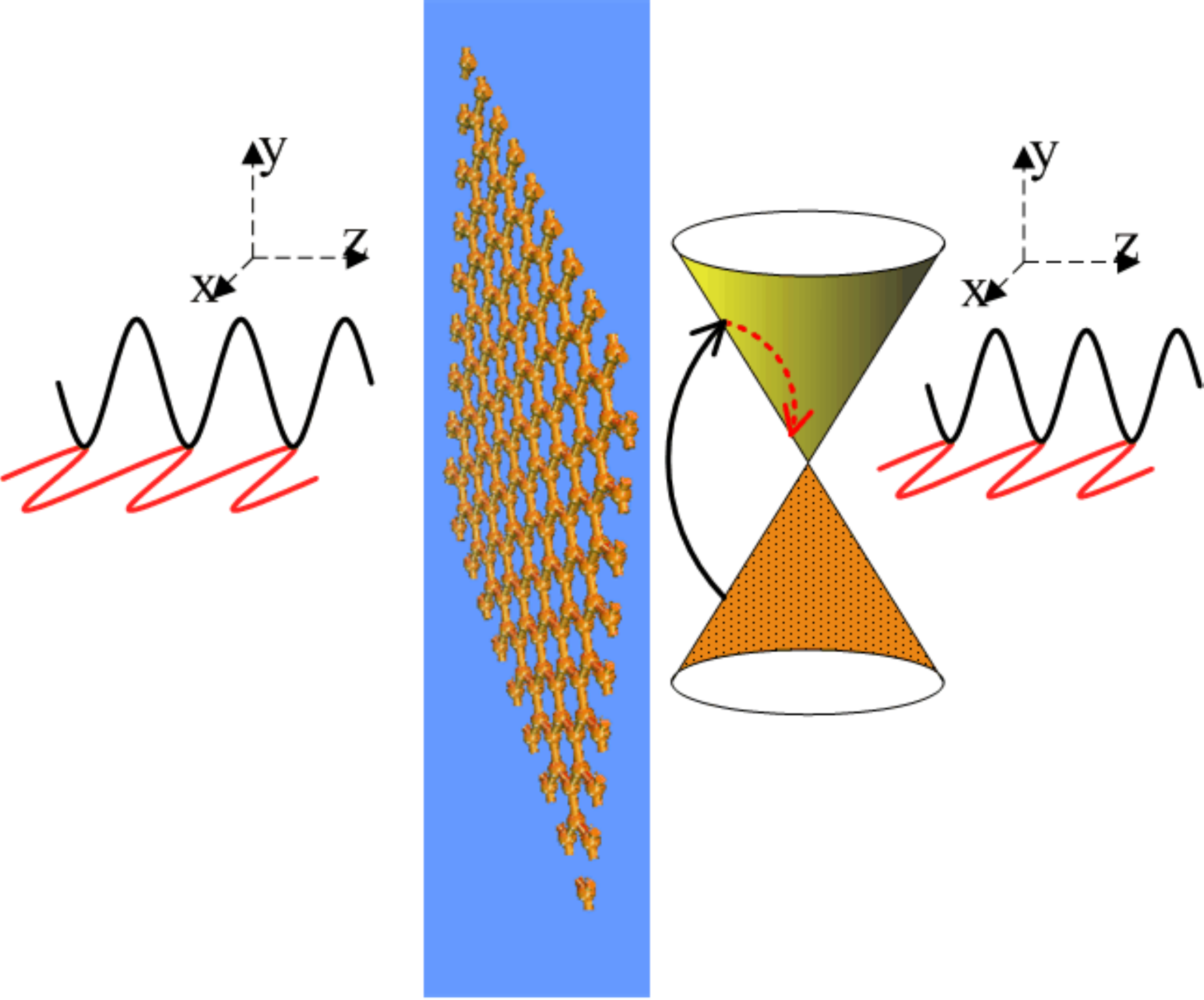}
\caption{Illustration of optical conductivity of graphene: Incident light normal to the graphene layer (x-z plane).}
\end{figure}

\section{Experimental studies}

We first experimentally investigated the polarization dependence of saturable absorption of the atomic layer
graphene using a setup as illustrated in Fig. 1. A stable mode locking L-band fiber laser was used as the light source.
 The output of the laser was first amplified through a commercial erbium-doped fiber amplifier (EDFA), and then separated into
 two orthogonally polarized components with an in-line polarization beam splitter (PBS), and perpendicularly incident to the
 2D atomic layer graphene. The saturable absorption feature of the graphene sample under the illumination of each of the orthogonally
 polarized light was measured, respectively. To reduce any artificial discrepancy caused by the fluctuation of the input power,
 output powers (with or without passing through the graphene sample) were detected simultaneously through two set of power meters.

  \begin{figure}
\centering
\includegraphics[width=8cm]{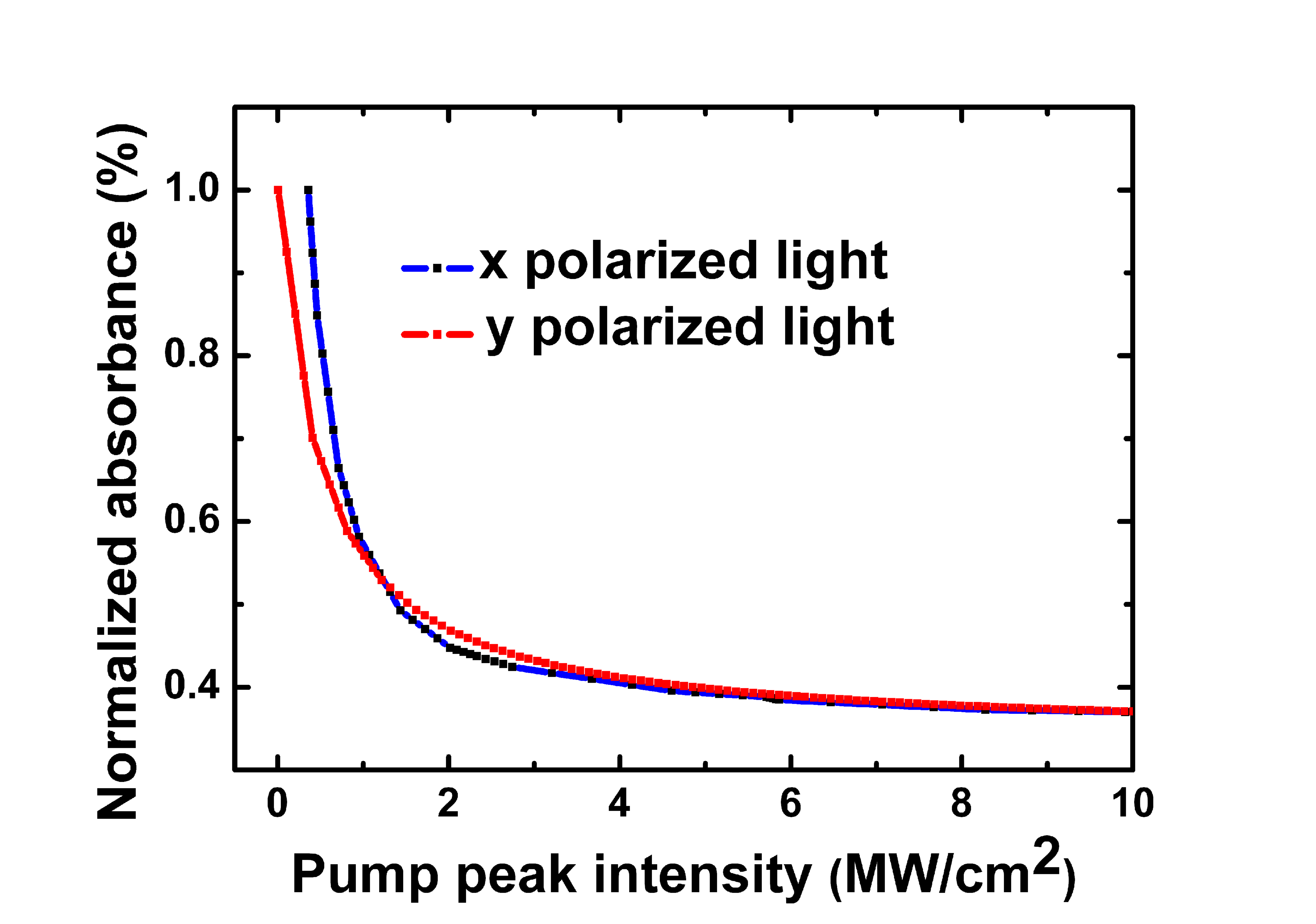}
\caption{Polarization resolved saturable absorption curve of graphene-based mode locker. }
\end{figure}

\subsection{Polarization dependent absorption of graphene }

 Fig. 2 shows the measured polarization dependence of the graphene saturable absorption.
 The same normalized modulation depth of ~66\%
 for both polarizations was observed. The measured saturable fluence for the two polarizations are ~0.71 MW cm$^-$$^2$ and ~0.70 MW cm$^-$$^2$, respectively, which is well within the measurement error range.

\subsection{Experimental Setup }

Using the atomic layer graphene as a mode locker, we further investigated the passive mode locking of an erbium-doped fiber
laser under various net cavity dispersions and the vector soliton formation in the laser. At first, an erbium-doped fiber laser
with net normal dispersion cavity as depicted schematically in Fig. 3 was constructed. The laser has a ring cavity that consists
of a segment of 5m Erbium-doped fiber (EDF) with erbium concentration of 2880 ppm and a group velocity dispersion
 (GVD) parameter of ?32 (ps/nm)/km, a total length of 9.0m single mode fiber (SMF) with a GVD parameter of   18 (ps/nm)/km,
 and ~118m dispersion compensation fiber (DCF) with a GVD parameter of ~2 (ps/nm)/km. The intra-cavity optical components such as
 the 10\% fiber coupler, wavelength division multiplexer (WDM) and optical isolator were carefully selected. They all have polarization dependent losses less than 0.1 dB. A polarization independent isolator was used to force the unidirectional operation of the ring, and an intra-cavity polarization controller (PC) was used to finely adjust the linear cavity birefringence. The graphene saturable absorber was inserted in the cavity through transferring a piece of free standing few layers graphene film onto the end facet of a fiber pigtail via Van Der Walls force. The laser was pumped by a high power Fiber Raman Laser source (KPS-BT2-RFL-1480-60-FA) of wavelength 1480 nm, whose maximum pump power can reach as high as 5 W.

 \begin{figure}
\centering
\includegraphics[width=8cm]{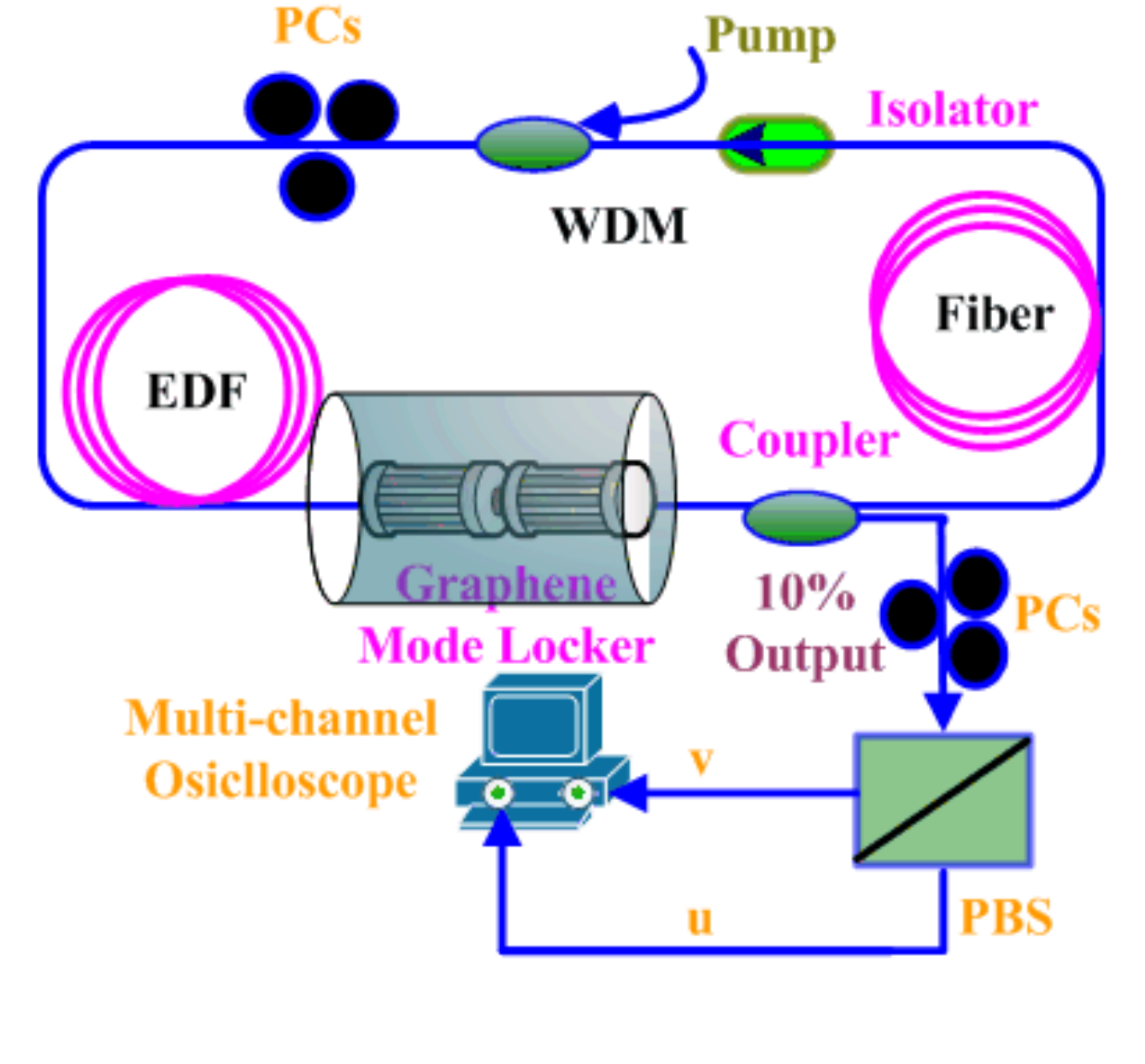}
\caption{Schematic of the soliton fiber laser. WDM: wavelength division multiplexer. EDF: erbium-doped fiber. PC: polarization controller. PBS: polarization beam splitter.  }
\end{figure}

The fiber laser has a typical dispersion-managed cavity configuration and its net cavity dispersion is ~0.3047 ps$^2$. Mode locking of the laser was achieved through slightly varying the orientation of the PC. Due to the nonlinear pulse propagation in the net normal dispersion cavity, the mode locked pulses were immediately automatically shaped into dissipative solitons. Fig. 4 shows a typical mode locked state of the laser. Fig. 4a is the mode locked pulse spectrum. It exhibits the characteristic sharp steep spectral edges of the dissipative solitons formed in the normal dispersion fiber lasers, indicating that the mode locked pulses have been shaped into dissipative solitons.  Using a 50 GHz high-speed oscilloscope (Tektronix CSA 8000) together with a 45 GHz photo-detector (New Focus 1014), we have measured the dissipative soliton pulse width and profile, as shown in Fig. 4b. The pulses have a width (FWHM) of ~111 ps. If the Sech2 pulse shape is assumed, it gives the pulse width of ~71 ps. The 3dB spectrum bandwidth of the pulses is ~7.18 nm, which gives a time-bandwidth product of ~63.7, indicating that the pulses are strongly chirped.

\begin{figure}
  \centering
  \subfigure[]{
    \label{fig:subfig:a} 
    \includegraphics[width=8cm]{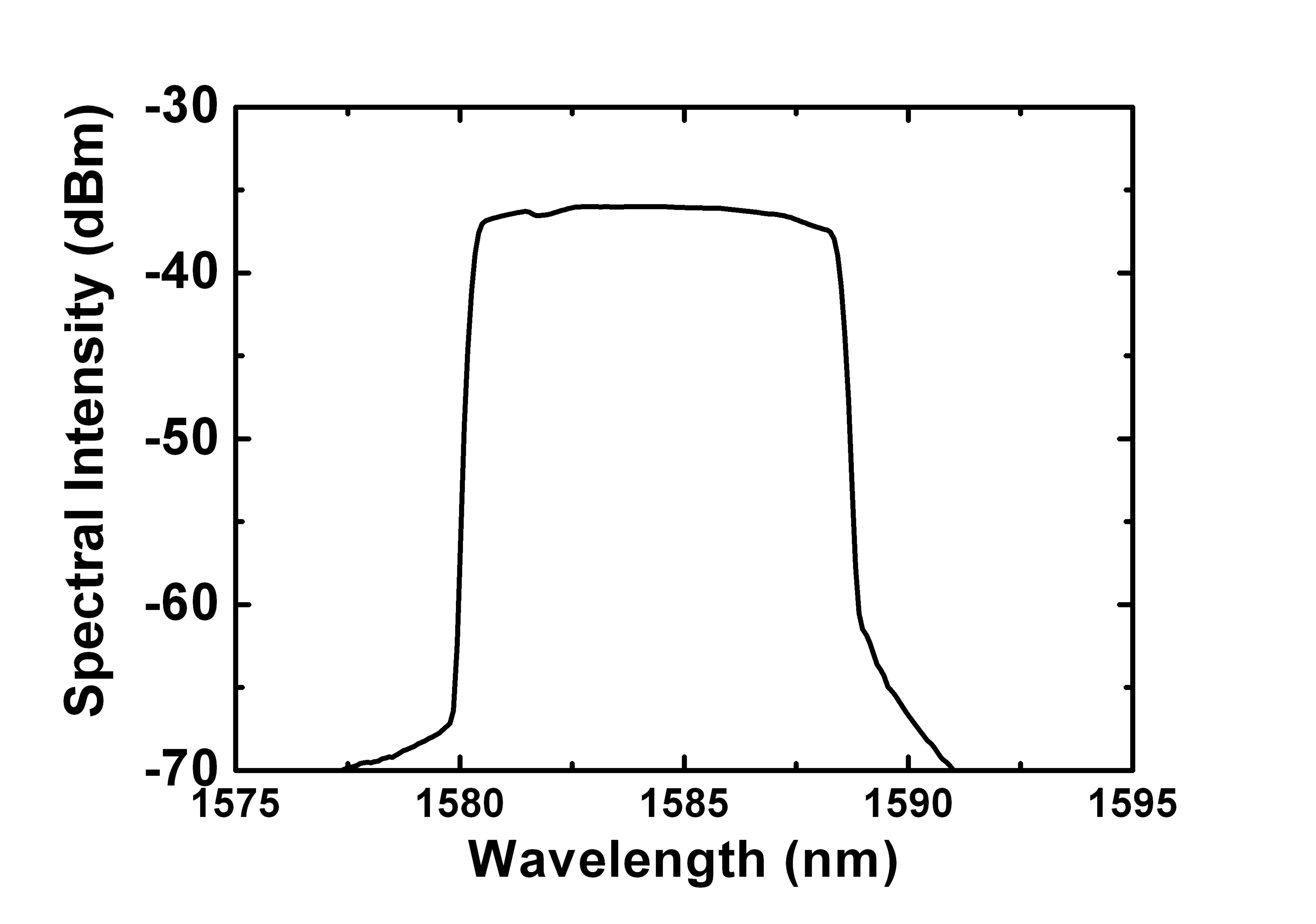}}
  \hspace{1in}
  \subfigure[]{
    \label{fig:subfig:b} 
    \includegraphics[width=8cm]{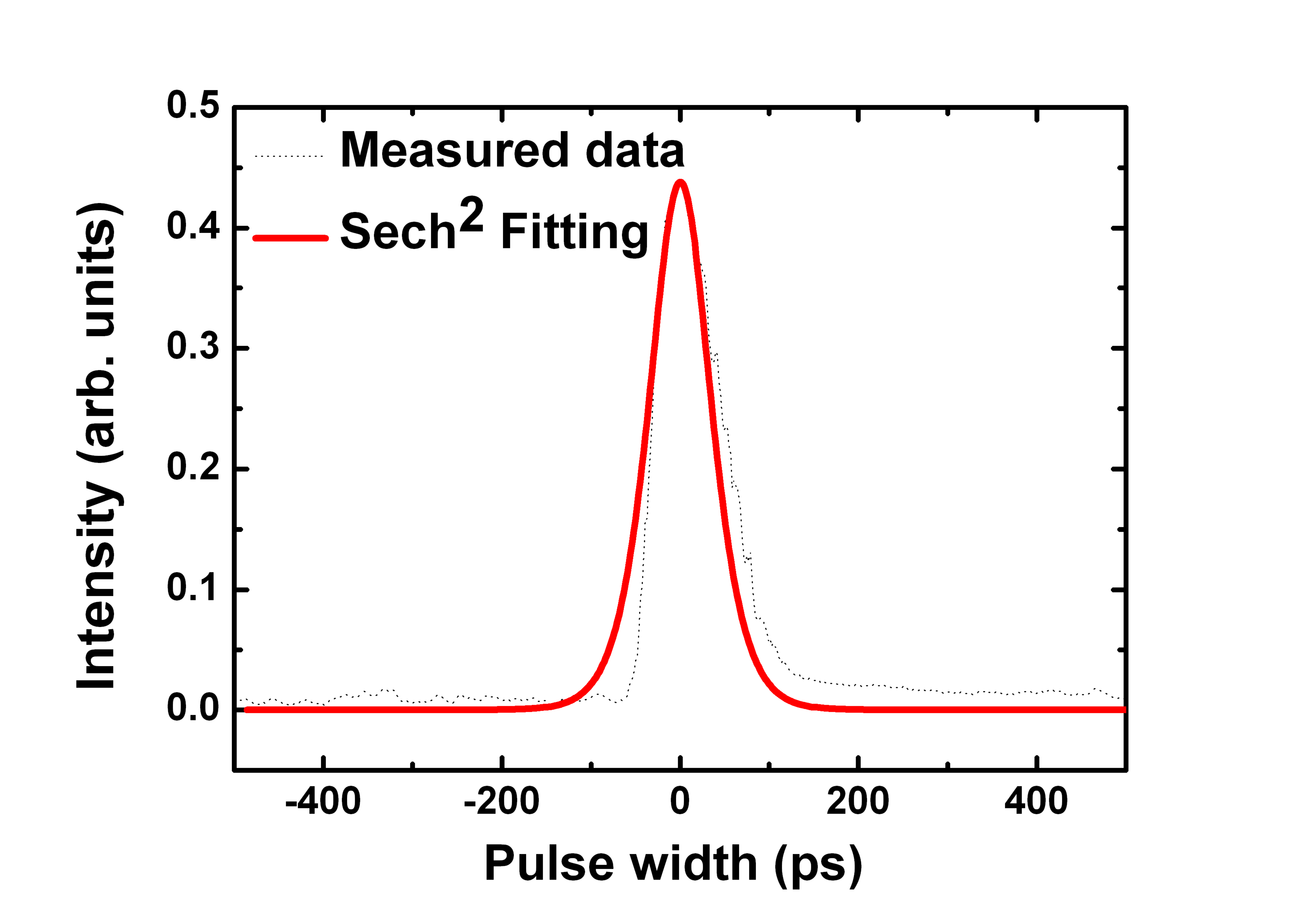}}
    \subfigure[]{
    \label{fig:subfig:b} 
    \includegraphics[width=8cm]{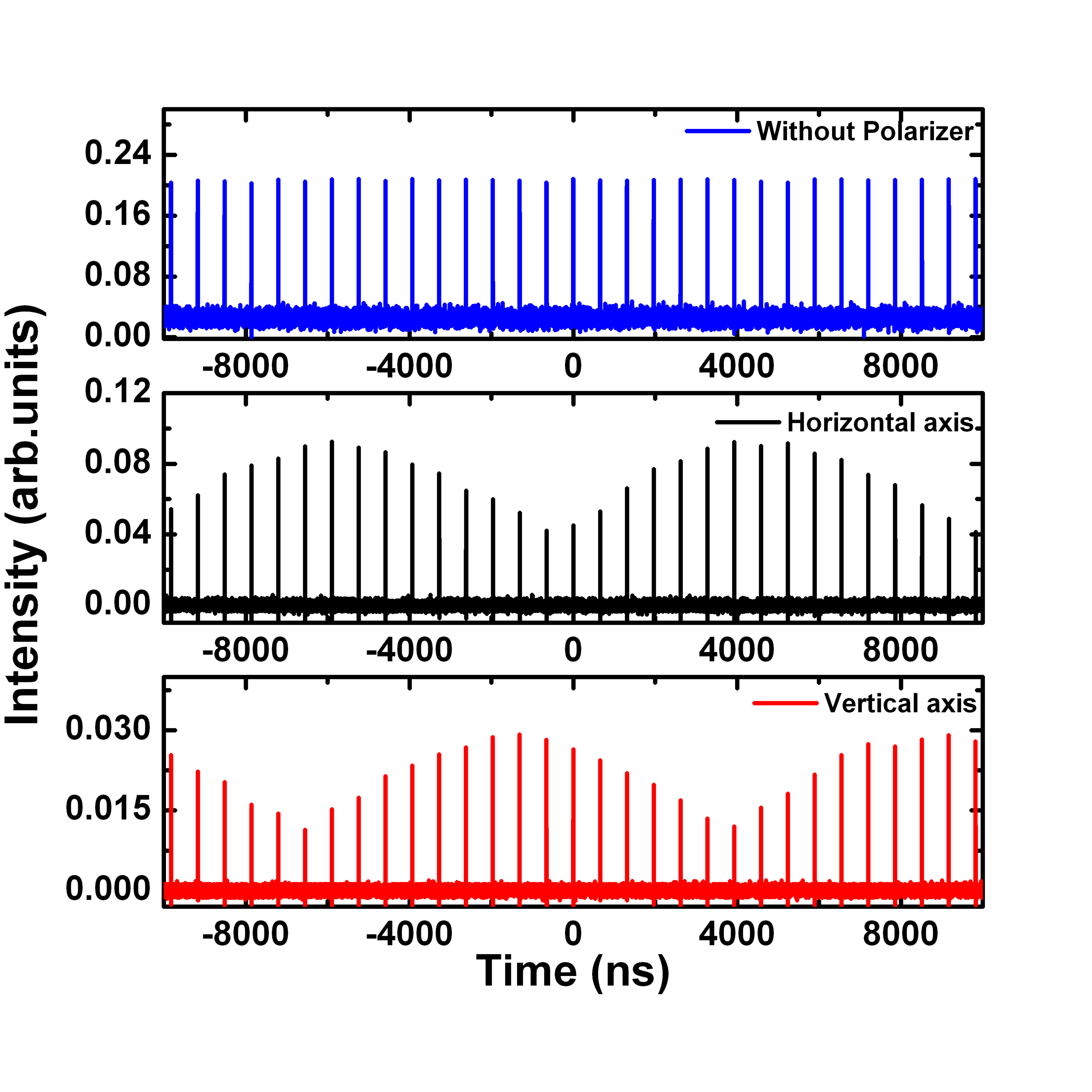}}
  \caption{Dissipative vector soliton operation of the fiber laser. (a) Optical spectrum measured. (b) Pulse profile measured with a high speed oscilloscope. (c) Polarization resolved oscilloscope traces: the polarization of the soliton is rotating in the cavity.}
  \label{fig:subfig} 
\end{figure}

\subsection{Dissipative vector solitons }

Experimentally we further identified that the formed dissipative solitons compose of two orthogonal polarization components, therefore, they are vector dissipative solitons. To experimentally resolve the two orthogonal polarization components of the solitons, an in-line polarization beam splitter (PBS) together with a PC were spliced to the output port of the fiber laser. The two orthogonally polarized outputs of the PBS were simultaneously measured with two identical 2GHz photo-detectors and monitored with a multi-channel oscilloscope (Agilent 54641A) and an optical spectrum analyzer (Ando AQ-6315B). Depending on the net cavity birefringence, either the polarization rotating or the polarization locked dissipative vector solitons were obtained in our experiments. Fig. 4c shows the polarization resolved measurement of a polarization rotation dissipative vector soliton experimentally obtained. Unlike the dissipative solitons observed in fiber lasers mode locked with the NPR technique, where the polarization of the solitons emitted by the laser is fixed, the polarization of the solitons varies from pulse to pulse. The polarization rotation feature of the dissipative soliton is clearly reflected in Fig. 4c. Without passing through a polarizer, the soliton pulse train has a uniform pulse height, while after passed through the PBS, the pulse height becomes periodically modulated. In particular, the periodic pulse height modulation for each of the orthogonal polarization directions is 90 degree out of phase. In Fig. 4c, the pulse height returns back to its original value after every 9.9 $\mu$s, which indicates that the polarization state of the DS rotates back to its original state after every 15 cavity roundtrips.

\begin{figure}
\centering
\includegraphics[width=8cm]{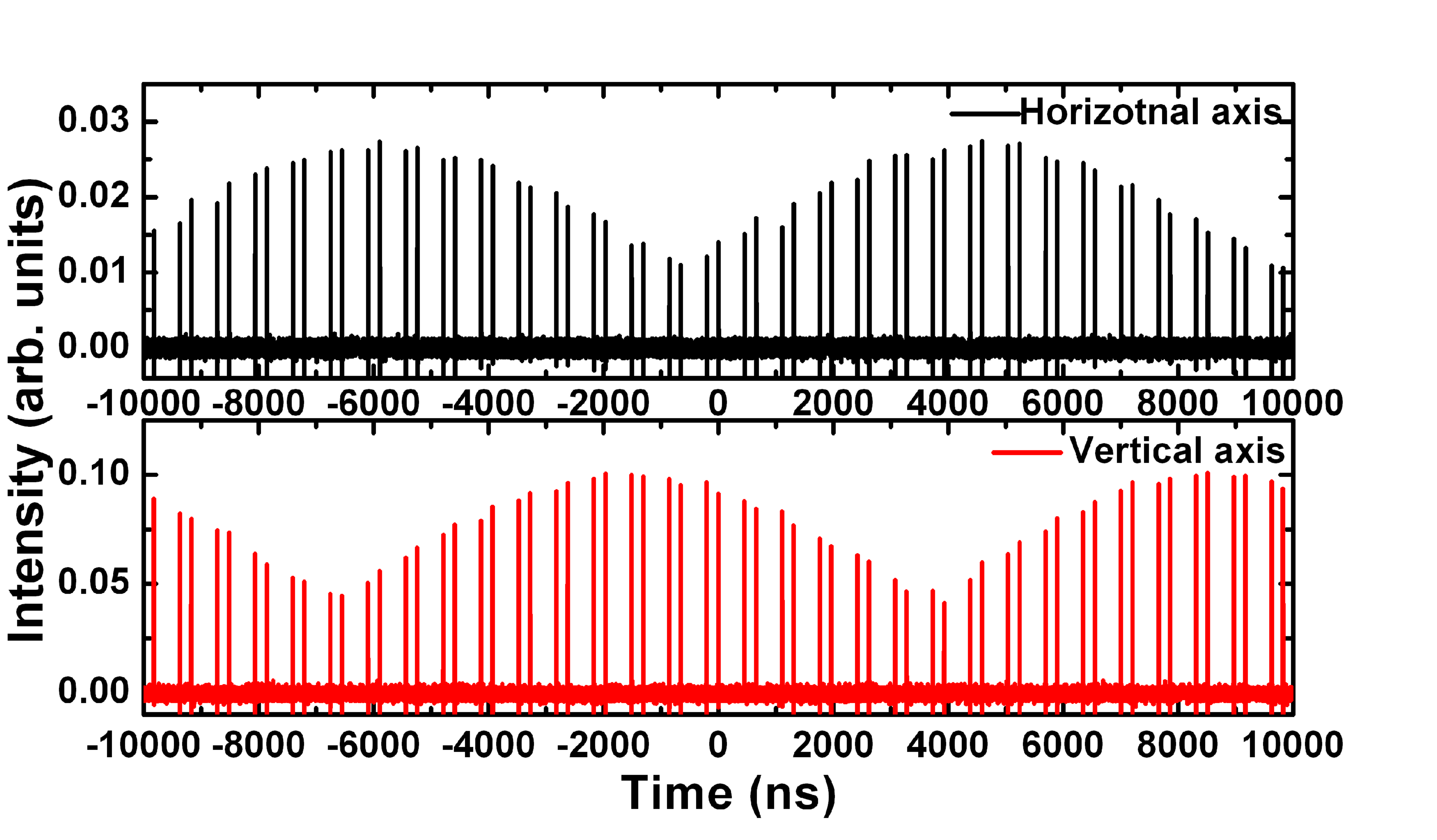}
\caption{Schematic of the soliton fiber laser. WDM: wavelength division multiplexer. EDF: erbium-doped fiber. PC: polarization controller. PBS: polarization beam splitter.  }
\end{figure}

In our experiment as the pump strength was increased to 1.5 W, a new dissipative soliton was formed in the cavity. Subsequently, a two DVSs operation state as shown in Fig. 5 was obtained. It was found that the polarization of the two DVSs rotated with the same speed in the cavity, despite of the fact that the polarization ellipse orientations of the two solitons are slightly different. We attribute the slight polarization orientation difference between the two solitons in the laser cavity to the result of gain competition between the vector solitons.

\begin{figure}
  \centering
  \subfigure[]{
    \label{fig:subfig:a} 
    \includegraphics[width=8cm]{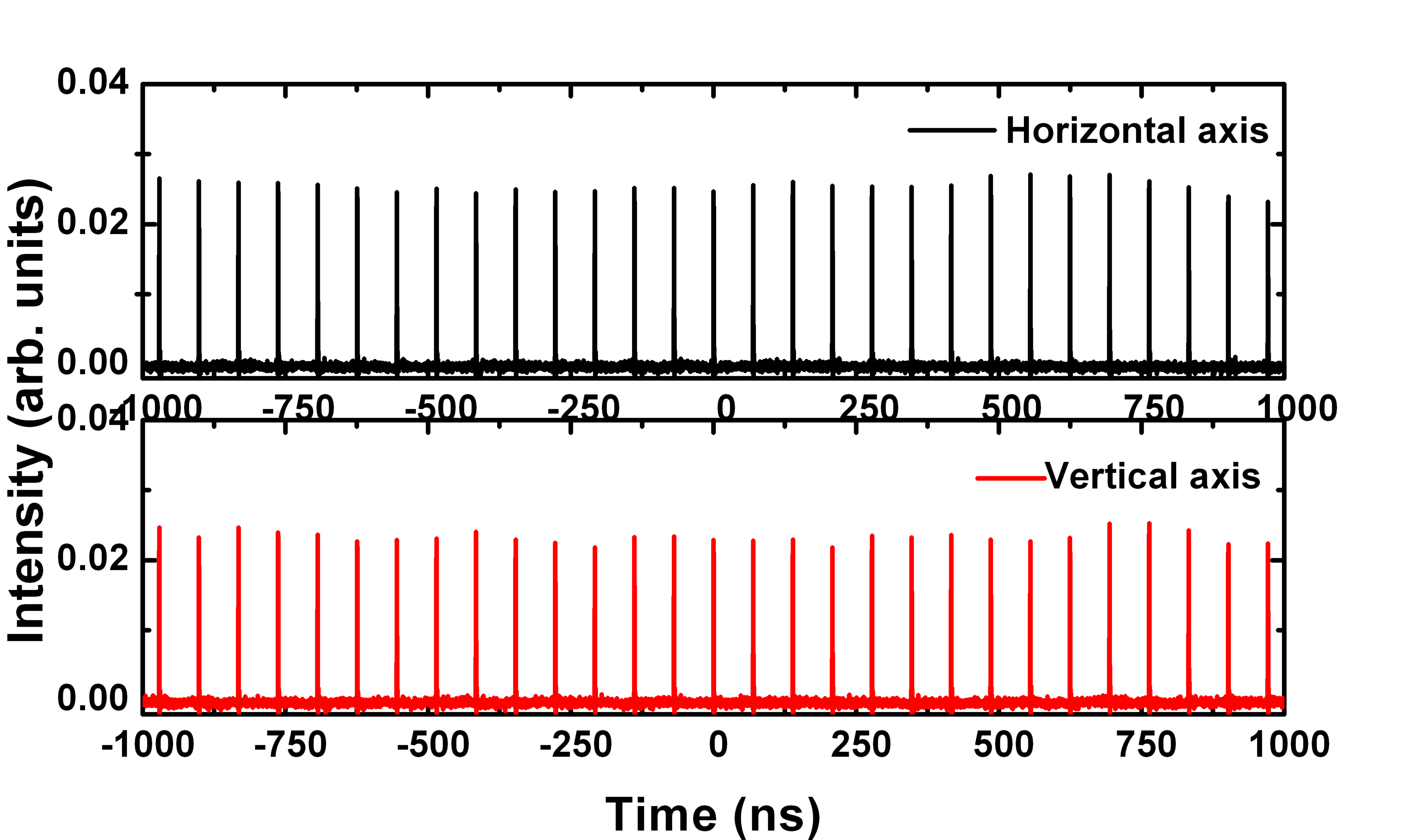}}
  \hspace{1in}
  \subfigure[]{
    \label{fig:subfig:b} 
    \includegraphics[width=8cm]{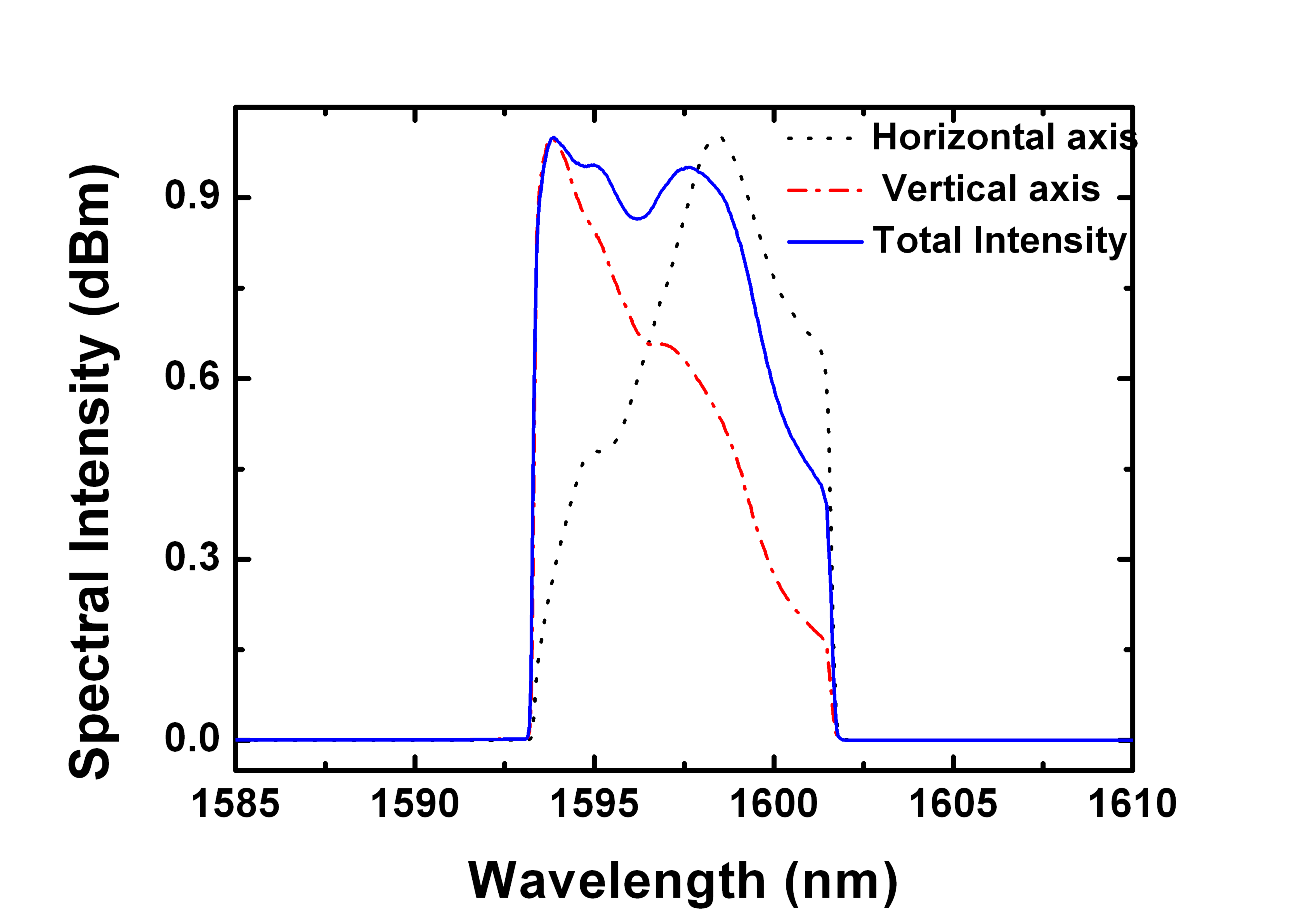}}
  \caption{Polarization locked dissipative soliton operation of the fiber laser. (a) Polarization resolved oscilloscope trace; (b) the corresponding optical spectra.}
  \label{fig:subfig} 
\end{figure}

Carefully adjusting the orientation of the intracavity PC, polarization locked dissipative vector solitons could also be obtained in our laser. Fig. 6 shows a state of the polarization locked DVS operation of the laser. In this case even measured after the extra cavity PBS, the two polarization-resolved pulse traces show uniform pulse trains. However, as the cavity birefringence was changed, the state then changed to a polarization rotation DVS state. For a phase locked DVS operation state, it is possible to measure the soliton spectra along the long and short polarization ellipse axes. They are shown in Fig. 6a. The two orthogonal polarization components have comparable spectral intensity and equal sharp edge to sharp edge spectral separation. However, carefully examining the fine structures of their spectra, it was found that they are obviously different. While the peak spectral position for one polarization component is at 1598.5 nm, the peak spectral position of the other is located at 1593.9 nm. The 3-dB spectral bandwidths of them are also different, one is 5.48 nm and the other is 5.58 nm. At first glance it is surprising that the peak spectral positions of the two orthogonal polarization components of a phase locked vector soliton could be different. However, considering that the vector soliton is a dissipative soliton, and the characteristics of the dissipative solitons are that they have large nonlinear frequency chirp and broad pulse width, this result might be possible.

\begin{figure}
  \centering
  \subfigure[]{
    \label{fig:subfig:a} 
    \includegraphics[width=7.5cm]{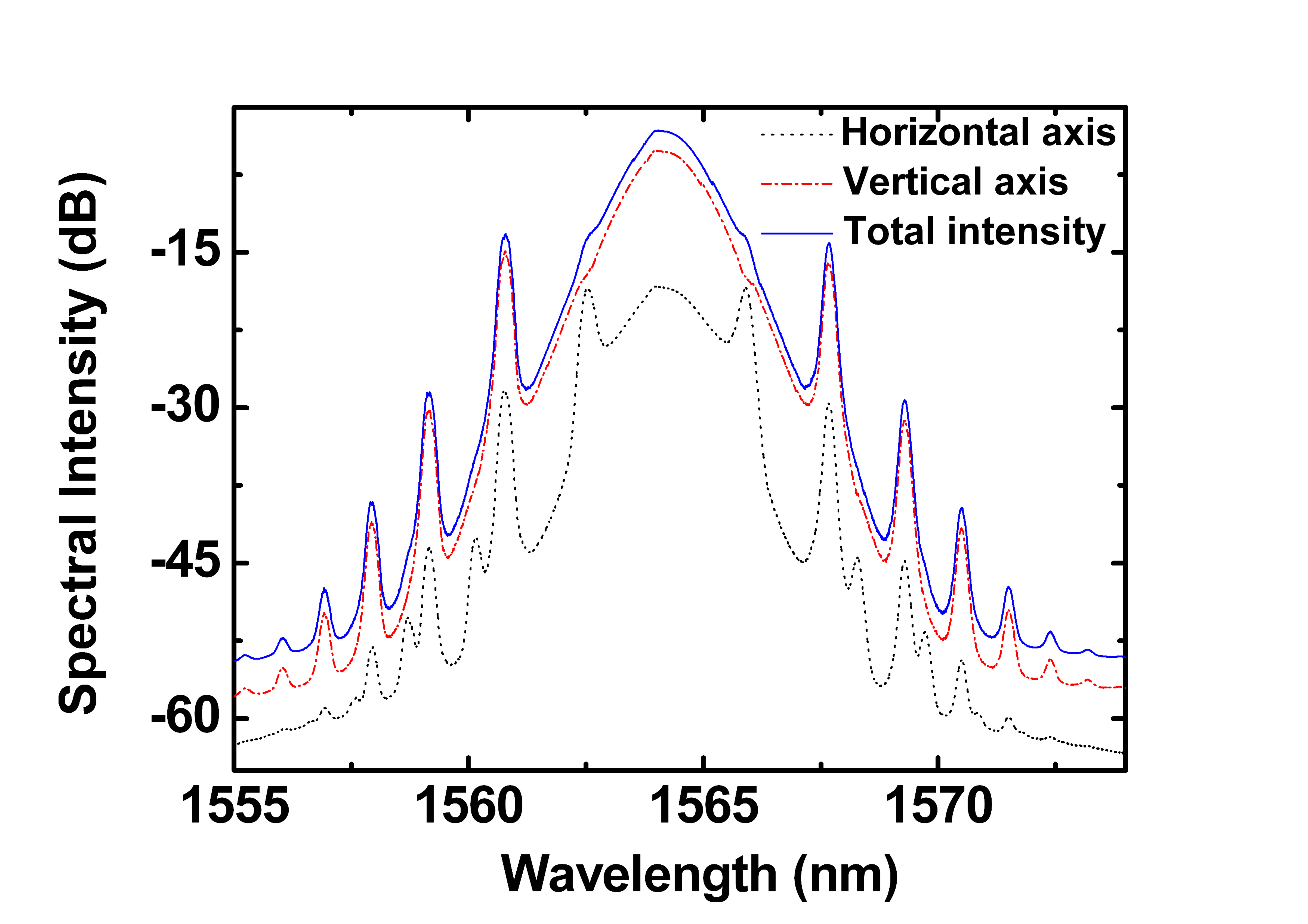}}
  \hspace{1in}
  \subfigure[]{
    \label{fig:subfig:b} 
    \includegraphics[width=7.5cm]{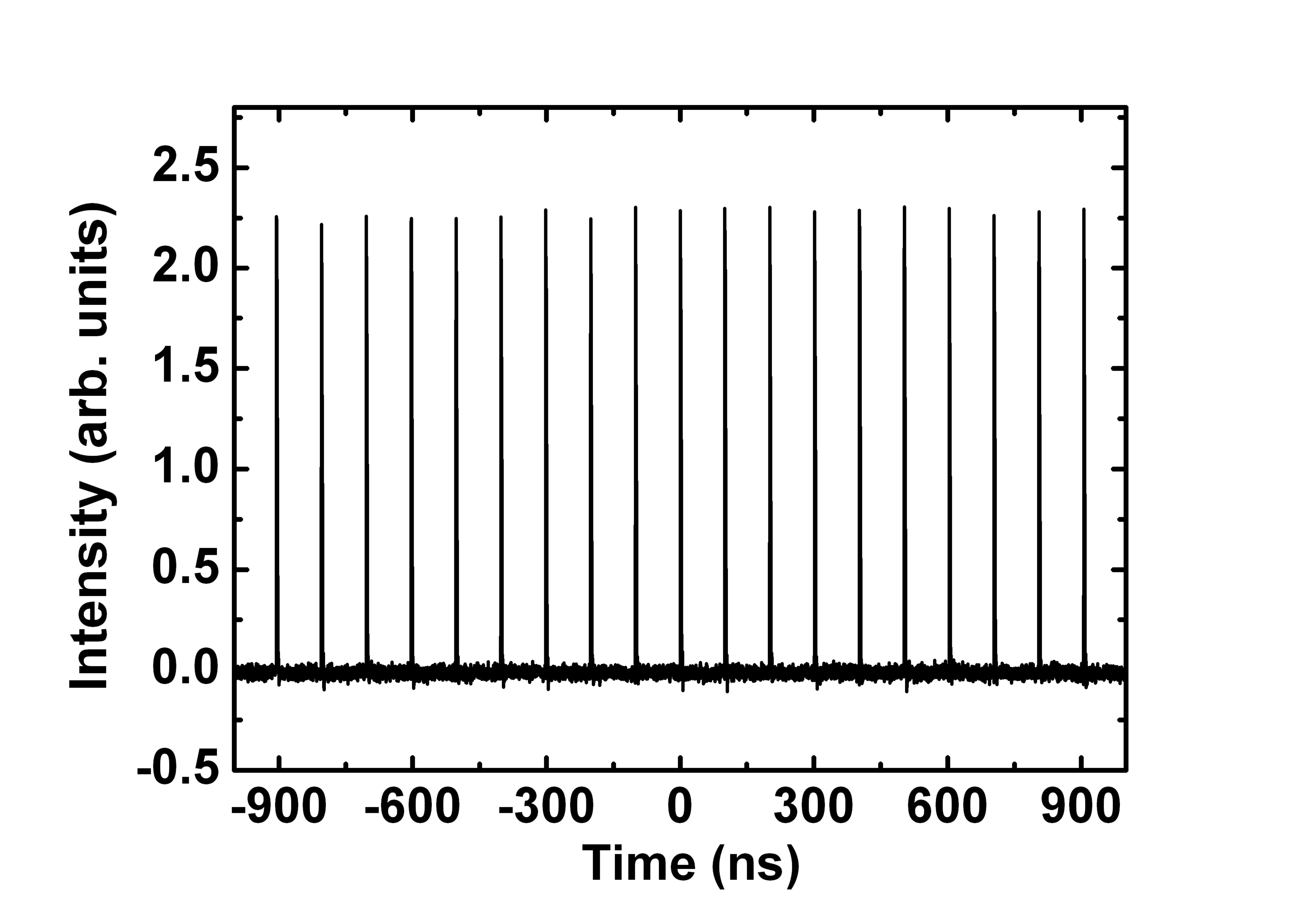}}
      \subfigure[]{
    \label{fig:subfig:b} 
    \includegraphics[width=7.5cm]{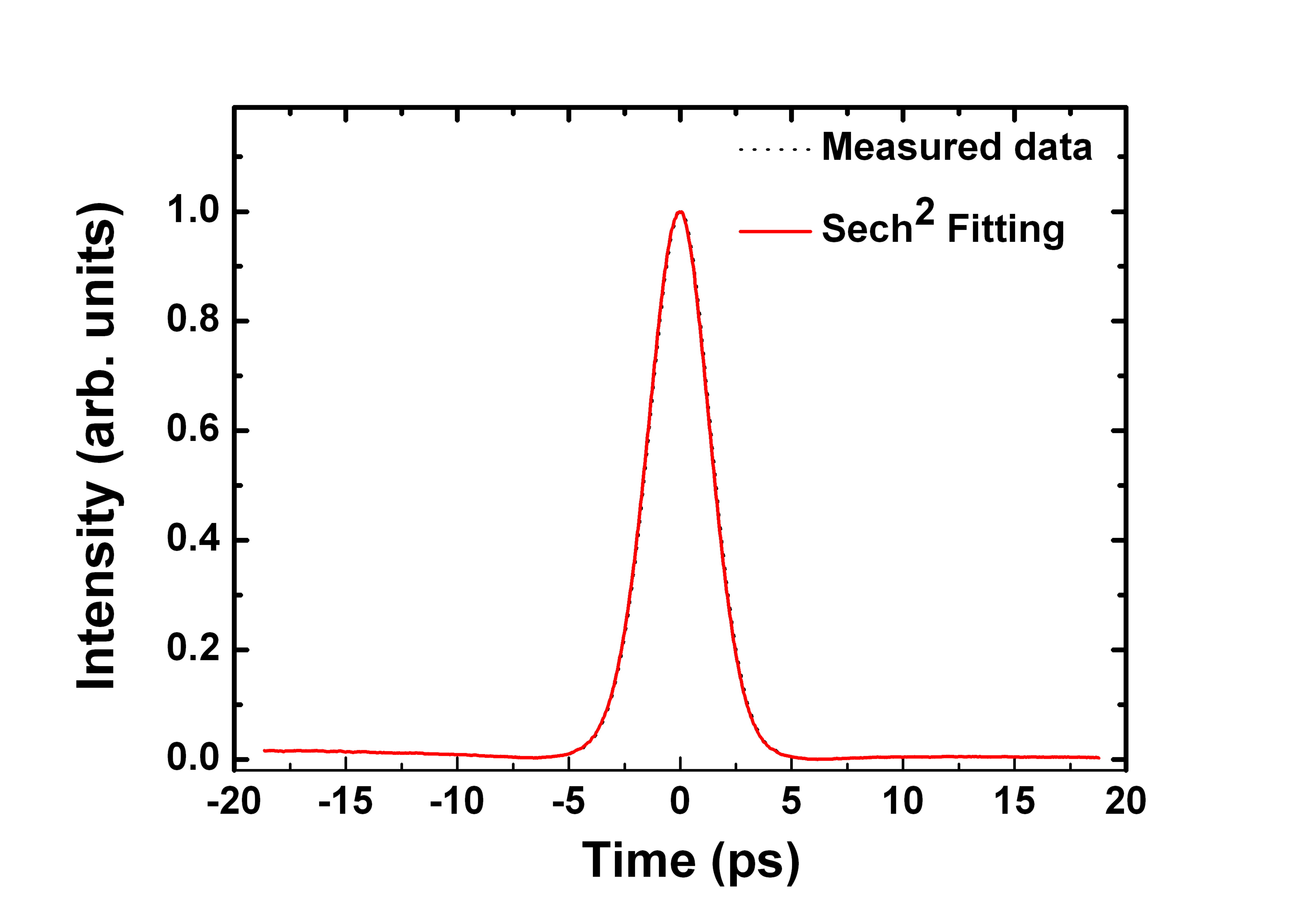}}
  \caption{A state of polarization locked NLSE vector soliton operation of the anomalous dispersion fiber laser. (a) Optical spectra of the vector soliton. (b) Soliton pulse train measured after an external cavity polarizer; (c) The corresponding autocorrelation trace.}
  \label{fig:subfig} 
\end{figure}

\subsection{Coherent energy exchange }

We had also experimentally investigated the vector soliton formation and dynamics in an all-anomalous-dispersion erbium-doped fiber laser mode locked with atomic layer graphene. The fiber laser has a similar cavity structure as shown in Fig. 2 but with all anomalous dispersion fibers. The ring laser cavity has a length of ~20m, comprising ~6 m EDF with group velocity dispersion (GVD) parameter of 10 (ps/nm)/km and ~ 14m SMF with GVD parameter of 18 (ps/nm)/km. Fig. 7a shows a typical mode locked state of the laser observed. Like the mode-locked pulses obtained in other anomalous dispersion cavity fiber lasers, the optical spectrum of the mode locked pulses exhibit the characteristic Kelly sidebands, a signature of the conventional NLSE soliton operation of the laser \cite{no22}. Fig. 7b is the measured oscilloscope trace after an external cavity polarizer and no intensity modulation was observed, indicating that this state is a polarization locked vector soliton operation state. The 3 dB bandwidth of the spectra is about 1.5 nm. Fig. 2c shows the measured autocorrelation trace of the solitons. It has a Sech2-profile with a FWHM width of about 3.1 ps, which, divided by the decorrelation factor of 1.54, corresponds to a pulse width of 2.1 ps. The time-bandwidth product (TBP) of the pulses is 0.37, showing that the solitons are nearly transform-limited. In Fig. 7a, we have also shown the optical spectra of the pulses along two orthogonal polarization directions, measured after the extra cavity PBS like in the dissipative soliton cases shown above. Obviously, the soliton spectra along the two orthogonal polarization directions have different spectral profiles and sidebands, showing that the solitons are vector solitons. Previously, we have experimentally investigated the vector NLSE soliton formation in a fiber laser mode locked with a SESAM \cite{no23}. It was found that a four-wave-mixing (FWM) type of new spectral sidebands could also form on the polarization resolved spectra of the vector solitons, and different from the Kelly sidebands whose positions on the soliton spectrum is almost invariant with the pumping strength and cavity birefringence, the positions of the FWM sidebands varies sensitively with the cavity birefringence. The formation of the FWM sidebands was due to the resonant energy exchange between the two orthogonal polarization components of the vector soliton. As it is an internal process of the vector soliton, on the total soliton spectrum these FWM spectral sidebands are invisible, but on the polarization resolved spectra they exhibit as the local peak-dip spectral spikes. In Fig. 7a the extra FWM sidebands are clear visible. The vector solitons shown in Fig. 7 are phase locked as identified from the polarization resolved oscilloscope traces. As the NLSE solitons formed in the laser are nearly transform-limited, the central wavelengths of the two phase locked solitons as well as their spectral sidebands have the same values.We had also experimentally investigated the vector soliton formation and dynamics in an all-anomalous-dispersion erbium-doped fiber laser mode locked with atomic layer graphene. The fiber laser has a similar cavity structure as shown in Fig. 2 but with all anomalous dispersion fibers. The ring laser cavity has a length of ~20m, comprising ~6 m EDF with group velocity dispersion (GVD) parameter of 10 (ps/nm)/km and ~ 14m SMF with GVD parameter of 18 (ps/nm)/km. Fig. 7a shows a typical mode locked state of the laser observed. Like the mode-locked pulses obtained in other anomalous dispersion cavity fiber lasers, the optical spectrum of the mode locked pulses exhibit the characteristic Kelly sidebands, a signature of the conventional NLSE soliton operation of the laser \cite{no22}. Fig. 7b is the measured oscilloscope trace after an external cavity polarizer and no intensity modulation was observed, indicating that this state is a polarization locked vector soliton operation state.

The 3 dB bandwidth of the spectra is about 1.5 nm. Fig. 2c shows the measured autocorrelation trace of the solitons. It has a Sech2-profile with a FWHM width of about 3.1 ps, which, divided by the decorrelation factor of 1.54, corresponds to a pulse width of 2.1 ps. The time-bandwidth product (TBP) of the pulses is 0.37, showing that the solitons are nearly transform-limited. In Fig. 7a, we have also shown the optical spectra of the pulses along two orthogonal polarization directions, measured after the extra cavity PBS like in the dissipative soliton cases shown above. Obviously, the soliton spectra along the two orthogonal polarization directions have different spectral profiles and sidebands, showing that the solitons are vector solitons. Previously, we have experimentally investigated the vector NLSE soliton formation in a fiber laser mode locked with a SESAM \cite{no23}. It was found that a four-wave-mixing (FWM) type of new spectral sidebands could also form on the polarization resolved spectra of the vector solitons, and different from the Kelly sidebands whose positions on the soliton spectrum is almost invariant with the pumping strength and cavity birefringence, the positions of the FWM sidebands varies sensitively with the cavity birefringence. The formation of the FWM sidebands was due to the resonant energy exchange between the two orthogonal polarization components of the vector soliton. As it is an internal process of the vector soliton, on the total soliton spectrum these FWM spectral sidebands are invisible, but on the polarization resolved spectra they exhibit as the local peak-dip spectral spikes. In Fig. 7a the extra FWM sidebands are clear visible. The vector solitons shown in Fig. 7 are phase locked as identified from the polarization resolved oscilloscope traces. As the NLSE solitons formed in the laser are nearly transform-limited, the central wavelengths of the two phase locked solitons as well as their spectral sidebands have the same values.

\begin{figure}
\centering
\includegraphics[width=8cm]{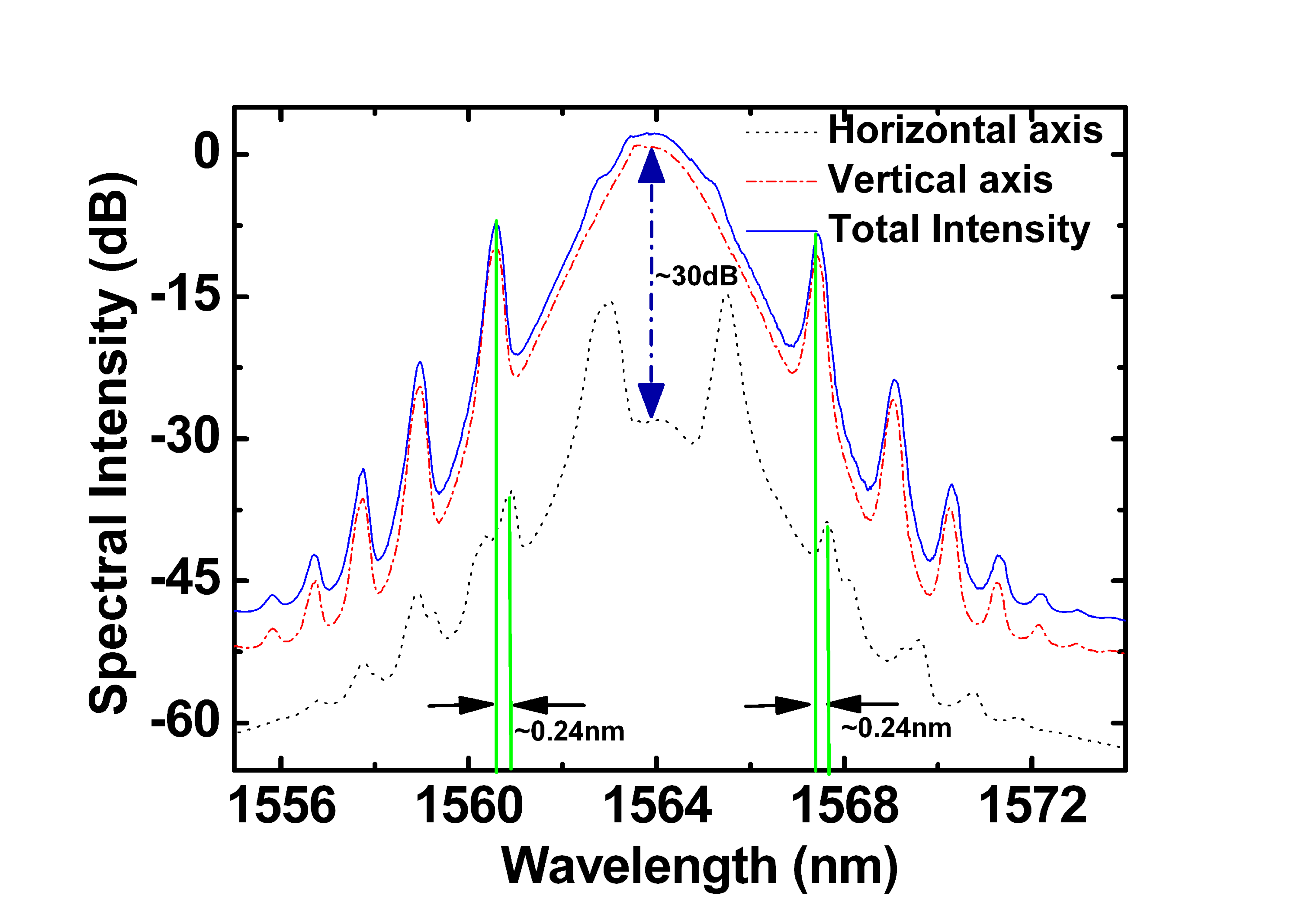}
\caption{A state of incoherently coupled vector soliton operation of the anomalous dispersion fiber laser. The weak soliton component is induced by the strong component through cross phase modulation.}
\end{figure}

\subsection{Induced soliotns }

Depending on the net cavity birefringence another vector soliton operation state of the laser as shown in Fig. 8 was also obtained. A strong soliton was formed along one principal polarization direction of the laser cavity, as evidenced by the clear Kelly sidebands. On the orthogonal polarization a weak soliton with different central wavelength was also formed. However, the Kelly sidebands of the weak soliton have different positions on the soliton spectra. As can be identified in Fig. 8, the first order Kelly sidebands of the strong soliton are located at 1560.57nm and 1567.39 nm, while those of the weak soliton are at the 1560.91nm and 1567.63nm, respectively. The spectral offset of the sidebands are the same ~0.24nm. We note that the FWM sidebands are now very strong on the weak soliton spectrum, but almost invisible on that of the strong soliton. The experimental result suggests that the weak soliton should be an induced soliton, formed due to the cross phase modulation of the strong soliton. Induced solitons were frequently observed in the fiber laser under relatively large net cavity birefringence \cite{no24}.

\subsection{Discussion and Prospects}

Our experiments have clearly shown that using atomic layer graphene as saturable absorber, various types of vector solitons can be formed in fiber lasers. The formation of the polarization rotation vector solitons unambiguously excluded the possibility that the mode locking of the laser could be due to the NPR effect. In our previous experiments on the NPR mode locked fiber lasers, we have noticed that the large residual polarization dependent losses of the cavity components, such as the fiber coupler or WDM, could cause NPR mode locking. To generate vector solitons one need to carefully select the cavity components. Moreover, recently saturable absorption of atomic layer graphene as laser mode locker has attracted considerable attention due to its super broadband absorption and fast recovery time \cite{no19,no20,no21}. Our current experimental results further show that under appropriate usage, the saturable absorption of graphene can also be light polarization insensitive.  We believe this feature of the graphene can also find interesting photonic applications.

\section{Conclusion}

In conclusion, we have experimentally investigated the vector soliton operation of erbium-doped fiber lasers mode locked with the atomic layer graphene. Our experimental results not only unambiguously confirmed the saturable absorption of the graphene and its application for passive fiber laser mode locking, but also showed that the saturable absorption of graphene is polarization insensitive when the light is incident perpendicular to the 2D atomic layer. Due to that the saturable modulation depth of graphene saturable absorber could be easily changed, we expected that new dynamics of vector solitons could be further observed in the graphene mode locked vector soliton fiber lasers.

\section{Acknowledgement}

Authors are indebted to Professor R. J. Knize of the United States Air Force Academy for useful discussions on graphene. The work is funded by the National Research Foundation of Singapore under the Contract No. NRF-G-CRP 2007-01. K. P. Loh wishes to acknowledge funding support from NRF-CRP Graphene Related Materials and Devices (R-143-000-360-281).

\section{Citations and References}\label{sec:endnotes}

\end{document}